\documentclass[useAMS,usenatbib]{mn2e} 
\usepackage{epsfig} \usepackage{amsmath} \usepackage{rotating}

\newcommand{\be}{\begin{equation}} \newcommand{\beq}{\begin{equation}}
\newcommand{\ba}{\begin{eqnarray}} \newcommand{\ee}{\end{equation}}
\newcommand{\eeq}{\end{equation}} \newcommand{\ea}{\end{eqnarray}}
  
\newcommand{\hs}{\hspace{1mm}} 
  
  \newcommand{\apj}{ApJ}
\newcommand{\aap}{A\&A} \newcommand{\apjl}{ApJL}
\newcommand{\mnras}{MNRAS} \newcommand{\aj}{AJ}
\newcommand{\apjs}{ApJS} 
\newcommand{\araa}{ARA\&A} 
\newcommand{\physrep}{Physics Reports}

\def\lsim{~\rlap{$<$}{\lower 1.0ex\hbox{$\sim$}}}

\def\gsim{~\rlap{$>$}{\lower 1.0ex\hbox{$\sim$}}}

\title[Ly$\alpha$ Acceleration of Galactic Supershells]{Acceleration
of Galactic Supershells by Ly$\alpha$ Radiation}

\author[Dijkstra, \& Loeb]{Mark Dijkstra
\thanks{E-mail:mdijkstr@cfa.harvard.edu} and Abraham Loeb \\ Astronomy
Department, Harvard University, 60 Garden Street, Cambridge, MA 02138,
USA\\}

\begin{document}

\date{\today} \pagerange{\pageref{firstpage}--\pageref{lastpage}}
\pubyear{2006}

\maketitle

\label{firstpage}
\begin{abstract}
Scattering of Ly$\alpha$ photons by neutral hydrogen gas in a single
outflowing 'supershell' around star forming regions often explains the
shape and offset of the observed Ly$\alpha$ emission line from
galaxies. We compute the radiation pressure that is exerted by this
scattered Ly$\alpha$ radiation on the outflowing material. We show
that for reasonable physical parameters, Ly$\alpha$ radiation pressure
alone can accelerate supershells to velocities in the range $v_{\rm
sh}=200$--$400$ km s$^{-1}$. These supershells possibly escape from
the gravitational potential well of their host galaxies and contribute
to the enrichment of the intergalactic medium. We compute the physical
properties of expanding supershells that are likely to be present in a
sample of known high-redshift ($z=2.7$--$5.0$) galaxies, under the
assumption that they are driven predominantly by Ly$\alpha$ radiation
pressure. We predict ranges of radii $r_{\rm sh}=0.1$--$10$ kpc, ages
$t_{\rm sh }=1$--$100$ Myr, and energies $E_{\rm
sh}=10^{53}$--$10^{55}$ ergs, which are in reasonable agreement with
the properties of local galactic supershells. Furthermore, we find
that the radius, $r_{\rm sh}$, of a Ly$\alpha$-driven supershell of
constant mass  depends uniquely on the intrinsic Ly$\alpha$ luminosity
of the galaxy, $L_{\alpha}$, the HI column density of the supershell,
$N_{\rm HI}$, and the shell speed, $v_{\rm sh}$, through the scaling
relation $r_{\rm sh}\propto L_{\alpha}/(N_{\rm HI}v_{\rm sh}^2)$. We
derive mass outflow rates in supershells that reach $\sim 10$--$100\%$
of the star formation rates of their host galaxies.
\end{abstract}

 
\section{Introduction}
\label{sec:intro}

The Ly$\alpha$ emission line of galaxies is often redshifted
relative to metal absorption lines
\citep[e.g.][]{Pettini01,Shapley03}. Scattering of Ly$\alpha$ photons
by neutral hydrogen atoms in an outflowing 'supershell' surrounding
the star forming regions can naturally explain this observation, as
well as the typically observed asymmetric spectral shape of the
Ly$\alpha$ emission line
\citep{L95,Lee98,T99,Ahn02,Ahn03,Mass03,Ahn04,Verhamme06,Verhamme08}. The
presence of an outflow may also be required to avoid complete
destruction\footnote{ Ly$\alpha$ may also avoid complete
destruction by interstellar dust when it is primarily locked up in
cold clumps embedded in a hot medium that is transparent to Ly$\alpha$
\citep{N91,H06}.} of the Ly$\alpha$ radiation by dust and to allow
its escape from the host galaxies
\citep[][]{Kunth98,Hayes08,Ostlin08,Atek08}.

Indeed, the existence of outflowing thin shells (with a thickness much
smaller than their radius) of neutral atomic hydrogen around HII
regions is confirmed by HI-observations of our own
\citep[][]{Heiles84} and other nearby galaxies
\citep[e.g.][]{Ryder95}. The largest of these shells, so-called
'supershells', have radii of $r_{\rm max}\sim 1$ kpc
\citep[e.g][]{Ryder95,M02,M06} and HI column densities in the range
$N_{\rm HI}\sim 10^{19}$--$10^{21}$ cm$^{-2}$
\citep[e.g.][]{L95,Kunth98,Ahn04,Verhamme08}. The existence of
larger extragalactic supershells has been inferred from spatially
extended ($\sim$ a few kpc) Ly$\alpha$ P-cygni profiles, that were
observed around local star burst galaxies \citep{Mass03}. These
'supershells' differ from more energetic 'superwinds'
\citep[e.g.][]{Heckman90,Martin05}, in which galactic-scale biconical
outflows break-out of the galaxies' interstellar medium with high
velocities (up to $\sim 10^3$ km s$^{-1}$, of which M82 is a classical
example). Supershells are thought to be generated by stellar winds or
supernovae explosions which sweep-up gas into a thin expanding neutral
shell \citep[see e.g.][for a review]{T88}. The back-scattering
mechanism attributes both the redshift and asymmetry of the Ly$\alpha$
line to the Doppler boost that Ly$\alpha$ photons undergo as they
scatter off the outflow on the far side of the galaxy back towards the
observer.

In this paper we show that the radiation pressure exerted by
backscattered Ly$\alpha$ radiation on the outflowing gas shell can
accelerate it to velocities that reach hundreds of km s$^{-1}$. Our
basic result can be illustrated through an order-of-magnitude
estimate. The net outward force exerted on a shell of
gas\footnote{The net momentum transfer rate per photon that
enters the shell is $\Delta {\bf
p}=\frac{h\nu_{\alpha}}{c}(1-\mu)${\bf e}$_r$, where {\bf e}$_r$
denotes a unit vector that is pointing radially outward. Furthermore,
$\mu={\bf k}_{\rm in}\cdot {\bf k}_{\rm out}$, where ${\bf k}_{\rm
in}$ (${\bf k}_{\rm out}$) denotes the propagation direction of the
Ly$\alpha$ photon as it enters (leaves) the shell. When averaged over
all directions, the total momentum transfer rate is
$\dot{N}_{\alpha}\frac{h\nu_{\alpha}}{c}${\bf e}$_r$, where
$\dot{N}_{\alpha}$ is the rate at which Ly$\alpha$ photons enter the
shell. This total net momentum transfer rate applies regardless
of the number of times each Ly$\alpha$ photon scatters inside the
shell, or regardless of whether the Ly$\alpha$ photon is absorbed by a
dust grain (and possibly re-emitted in the infrared).} by
backscattered Ly$\alpha$ radiation of luminosity $L_{{\rm BS},\alpha}$
is ${d(m_{\rm sh}v_{\rm sh})}/{dt}=L_{{\rm BS},\alpha}/c$. Here,
$m_{\rm sh}$ is the total mass in the shell, $v_{\rm sh}$ is the shell
velocity and $c$ is the speed of light. If we assume that
backscattering occurs over a timescale $t_{\rm BS}$, and the shell
mass is constant in time, then the net velocity gain by the shell is
$\Delta v_{\rm sh}\sim 250$ km s$^{-1}$ $(m_{\rm sh}/10^7 \hs
M_{\odot})^{-1}(t_{\rm BS}/50 \hs{\rm Myr})(L_{{\rm
BS},\alpha}/10^{43} \hs {\rm erg/s})$. The adopted $m_{\rm sh}$ value
corresponds to a thin spherical HI shell with a column density $N_{\rm
HI}=10^{20}$ cm$^{-2}$ and radius $r=1$ kpc.  The fiducial values of
$L_{{\rm BS},\alpha}$ and $t_{\rm BS}$ were chosen to yield the
typical observed shell speed and expected shell lifetime in
Lyman-break galaxies \citep[LBGs; see e.g.][]{Shapley03,Verhamme08}.
Here, and throughout this paper, we assume that the supershells are
mostly neutral. This assumption is based on 21-cm observations of
local supershells which indicate that these are very thin (thickness
$\ll$ their radius, e.g Heiles 1984) and therefore dense.  At these
high densities one expects free electrons and protons to recombine
efficiently and make the gas significantly neutral.  However,
\citet{P00} argue that, based on the observed reddening of the UV
continuum spectrum, either the dust-to-gas ratio in the supershell of
LBG MS1512-cB58 is higher than the galactic value, or alternatively
that the HI only makes up $f_{\rm HI}\sim 14-33\%$ of the total gas
mass of the shell, with the remainder of the gas being molecular or
ionized. We discuss how this latter possibility affects our results in
\S~\ref{sec:conc}.

This paper is a closely related to another paper in which detailed
Monte-Carlo Ly$\alpha$ radiative transfer calculations are performed
to assess Ly$\alpha$ radiation pressure in a more general context
(Dijkstra \& Loeb 2008, hereafter Paper I). These radiative transfer
calculations illustrate clearly that Ly$\alpha$ radiation pressure may
be important in driving outflows in various environments. Importantly,
the pressure exerted by Ly$\alpha$ radiation alone can exceed the
maximum possible pressure exerted by continuum radiation through dust
opacity (see Paper I), which in turn can exceed the total kinetic
pressure exerted by supernova ejecta \citep{Murray05}. This implies
that Ly$\alpha$ radiation pressure may thus in some cases provide the
dominant source of pressure on neutral hydrogen gas in the
interstellar medium (ISM).

The goals of this paper are ({\it i}) to gauge the general importance
of Ly$\alpha$ radiation  in the particular context of outflowing
galactic supershells, and ({\it ii}) to explore the related observable
properties of Ly$\alpha$-driven supershells.
In \S~\ref{sec:result} we compute the time evolution of the velocity
of an expanding shell of neutral gas that encloses a central
Ly$\alpha$ source. We consider a range of models, which are
subsequently applied to known Ly$\alpha$ emitting galaxies at high
redshifts. Finally, we discuss our results and their implications in
\S~\ref{sec:conc}. The parameters for the background cosmology used
throughout our discussion are
$(\Omega_m,\Omega_{\Lambda},\Omega_b,h)=(0.27,0.73,0.042,0.70)$
\citep{Dunkley08,Komatsu08}.
 
\section{Pressure from Backscattered Ly$\alpha$ Radiation}
\label{sec:result}

As mentioned in \S~\ref{sec:intro}, the net momentum transfer rate
from the Ly$\alpha$ radiation field to the shell is given by
\begin{equation}
\frac{d(m_{\rm sh}v_{\rm sh})}{dt}=m_{\rm sh}\frac{dv_{\rm
sh}}{dt}=f_{\rm scat}(v_{\rm sh},N_{\rm HI})\frac{L_{\alpha}}{c},
\label{eq:diff}
\end{equation} where $L_{\alpha}$ is the total Ly$\alpha$ luminosity of the galaxy, and $f_{\rm scat}(v_{\rm sh},N_{\rm HI})$ denotes the fraction of Ly$\alpha$ photons that are scattered in the shell. The latter depends on various factors, including the column density of neutral hydrogen atoms in the shell, the
intrinsic Ly$\alpha$ spectrum (i.e. the emitted spectrum {\it prior}
to scattering in the shell), as well as the shell velocity. For
example, as $v_{\rm sh}$ increases, atoms in the shell interact with
Ly$\alpha$ photons that are farther in the wing of the line profile,
and $f_{\rm scat}$ diminishes. Eq.~\ref{eq:diff} assumes that the HI
shell mass remains constant in time. This assumption is motivated by
the observation that supershells are well defined shells of HI gas
that are physically separate from the central Ly$\alpha$ source. These
outflows differ from steady-state outflows (i.e. outflows in which
mass is ejected at a constant rate $\dot{M}$) that occur around
late-type stars \citep[e.g.][]{salpeter74,Ivezic95}. In reality, an HI
supershell is not likely to maintain a constant mass as it sweeps
through the ISM, but its evolution depends on the detailed
hydrodynamics of the shell as it propagates through the ISM, which in
turn depends on the assumed properties of the ISM. In this first
study, we focus on simplified models that gauge the general importance
of Ly$\alpha$ radiation pressure on the dynamics of galactic
supershells.

Furthermore, our calculation of the parameter $f_{\rm scat}$
conservatively ignores that possibility of 'trapped' Ly$\alpha$
photons by the expanding supershell, which can boost the
radiation pressure considerably. If radiation trapping were included,
then the parameter $f_{\rm scat}$ would equal the 'force-multiplier'
$M_F$, which was shown to be much greater than unity at low $v_{\rm
sh}$ in Paper I (see \S~\ref{sec:conc} for a more detailed
discussion). We adopt a conservative approach because the
calculation of $M_F$ in Paper I assumed that the expanding shell
contained no dust. Under realistic circumstances, the existence of
dust inside the supershell will suppress the number of times the
Ly$\alpha$ photons can 'bounce' back and forth between opposite sites
of the shell, which leads to a reduction in the value of
$M_F$. However, provided that the dust is located inside (or exterior
to) the HI shell, the true value of $M_F$ is always larger than the
$f_{\rm scat}$ adopted in the paper (see \S~\ref{sec:conc}).

\subsection{Calculation of $f_{\rm scat}$}
\label{sec:fscat}

\begin{figure}
\vbox{\centerline{\epsfig{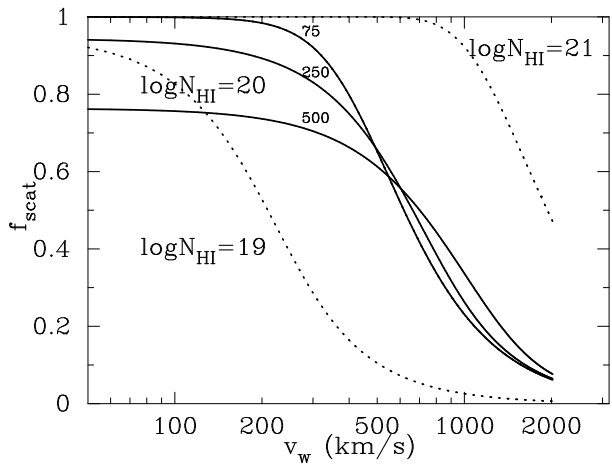}}}
\caption[]{The fraction of scattered Ly$\alpha$ photons $f_{\rm scat}$
as a function of shell velocity $v_{\rm sh}$ for a range of column
densities $N_{\rm HI}$ (in cm$^{-2}$, for $\sigma=75$ km
s$^{-1}$) and emitted line widths quantified by $\sigma$ (labeled by
75, 250 and 500 km s$^{-1}$ for $\log N_{\rm HI}=20$). Higher $N_{\rm
HI}$ values require larger shell speeds in order for the Ly$\alpha$
photons to propagate through the shell unobscured, because as $N_{\rm
HI}$ increases the photons need to be farther in the wing of the line
profile (in the frame of the shell) for them not to
scatter. Furthermore, a larger intrinsic line width, $\sigma$, serves
to flatten the dependence of $f_{\rm scat}$ on the shell speed (see
text).}
\label{fig:fw}
\end{figure} 
We compute $f_{\rm scat}(v_{\rm sh},N_{\rm HI})$ under the
conventional set of assumptions: ({\it i}) the emitted Ly$\alpha$
spectrum prior to scattering is assumed to be a Gaussian with a
velocity width $\sigma=v_{\rm circ}$ \citep{Santos04,igm}, where
$v_{\rm circ}$ is the circular virial velocity of the host dark matter
halo; and ({\it ii}) the outflow is modeled as {\it a single expanding
thin shell of gas} with a column density $N_{\rm HI}$ \citep[as
in][]{Ahn03,Ahn04,Verhamme06,Verhamme08}.  We can then write
\begin{equation}
f_{\rm scat}(v_{\rm sh},N_{\rm
HI})=\frac{1}{\sqrt{2\pi}\sigma_{x}}\int dx\hs{\rm
e}^{-x^2/2\sigma_{x}^2}\hs e^{-N_{\rm HI}\sigma_{\alpha}(x-x_{\rm
sh})},
\label{eq:fscat}
\end{equation} 
where frequency is denoted by the dimensionless variable $x\equiv
(\nu-\nu_0)/\Delta \nu_D$, with $\Delta \nu_D=v_{th}\nu_0/c$ and
$v_{th}$ being the thermal velocity of the hydrogen atoms in the gas
given by $v_{th}=\sqrt{2k_B T/m_p}$, $k_B$ is the Boltzmann constant,
$T=10^4$ K is the gas temperature, $m_p$ is the proton mass,
$\nu_0=2.47 \times 10^{15}$ Hz is the Ly$\alpha$ resonance frequency,
$\sigma_x=\sigma/v_{\rm th}$, and $x_{\rm sh}=v_{\rm sh}/v_{\rm
th}$. In Figure~\ref{fig:fw} we show $f_{\rm scat}$ as a function of
$v_{\rm sh}$ for a range of column densities $N_{\rm HI}$ (for
$\sigma=75$ km s$^{-1}$) and emitted line widths $\sigma$ (for $N_{\rm
HI}=10^{20}$ cm$^{-2}$).

As illustrated in Figure~\ref{fig:fw}, $f_{\rm scat}$ increases with
increasing $N_{\rm HI}$, because photons need to be farther in the
wing of the line profile (in the frame of the shell) for them not to
scatter. For example, Figure~\ref{fig:fw} shows that $\sim 50\%$ of
the Ly$\alpha$ photons are scattered when the shell speed is $\sim
200$ km s$^{-1}$ for $N_{\rm HI}=10^{19}$ cm$^{-2}$, whereas the same
fraction is reached when the shell speed is $\sim 600$ (2000) km
s$^{-1}$ for $N_{\rm HI}=10^{20}$ ($N_{\rm HI}=10^{21}$) cm$^{-2}$.
Furthermore, increasing the intrinsic line width serves to flatten the
dependence of $f_{\rm scat}$ on the shell speed, because at large
$\sigma$ and low shell speeds there is a substantial fraction of
photons far in the wing of the line profile.

\subsection{The Time Evolution of the Shell Speed: General Calculations}
\label{sec:vshell}

\begin{figure*}
\vbox{\centerline{\epsfig{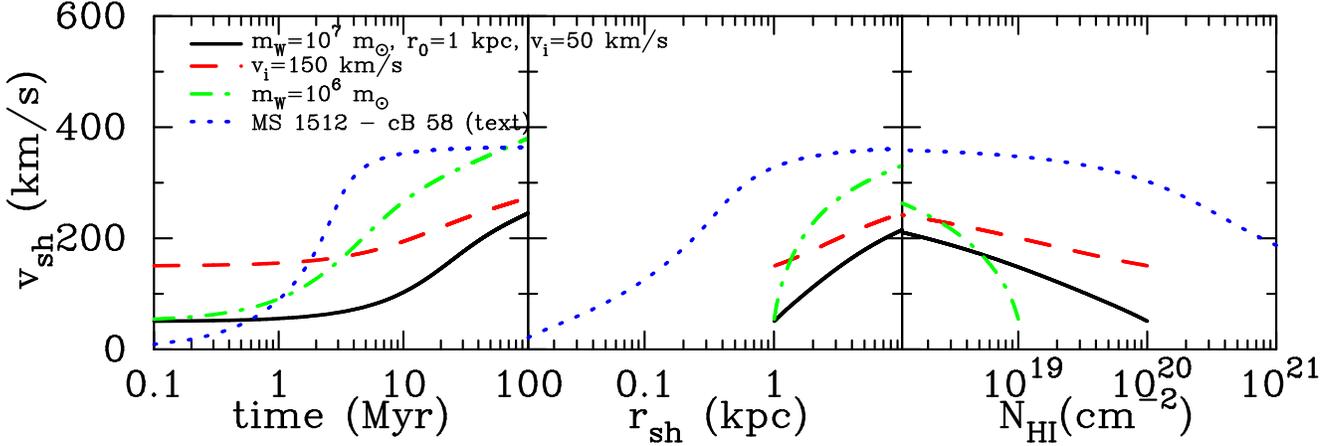}}}
\caption[]{{\it Left panel:} Time evolution of the speed of an
outflowing shell of neutral hydrogen that is accelerated by Ly$\alpha$
radiation from a star-forming galaxy. The {\it solid line} represents
our fiducial model (see text and Table~\ref{table:param}), in which
Ly$\alpha$ scattering accelerates the shell to $\sim 100$ km s$^{-1}$
in 10 Myr and to $\sim 230$ km s$^{-1}$ in 100 Myr. Other curves
represent models in which one of the model parameters was modified
(except for the specific example illustrated by the {\it blue dotted
line}; see \S~\ref{sec:modelpar}).  The curves demonstrate that
scattering of Ly$\alpha$ photons off an expanding shell can accelerate
the shell to speeds that reach a few hundred km s$^{-1}$.  {\it
Central \& Right panels} shows the shell velocity as a function of its
radius \& column density, respectively. In all examples except the
dotted line, the shell starts at $r=1$ kpc with a modest column
density. In the model of LBG MS1512-cB58 (dotted line), the shell
starts at $r=10^{-2}$ kpc with a higher column density that declines
to the observed value as the shell expands (see text).}
\label{fig:speed}
\end{figure*} 
Given $f_{\rm scat}(v_{\rm sh},N_{\rm HI})$ we next use
equation~(\ref{eq:diff}) to compute the evolution of the shell
position ($r_{\rm sh}$) and speed ($v_{\rm sh}$) as follows:

\noindent
{\bf 1.} The initial location and shell speed are denoted by $r_{\rm
sh,i}$ and $v_{\rm sh,i}$. If the outflowing shell covers $4\pi$
steradians of the sky surrounding the Ly$\alpha$ emitting region
\citep[see][for more detailed discussions]{Ahn04,Verhamme08}, and if
all the shell material is in neutral atomic hydrogen gas then $m_{\rm
sh}=4\pi r_{\rm sh,i}^2 N_{\rm HI,i}m_p$, where $r$ is the radius of
the shell, $m_p=1.6\times 10^{-24}$ g is the proton mass, and $N_{\rm
HI,i}$ is the initial column density of neutral hydrogen atoms.

\noindent
{\bf 2.} We combine equations~(\ref{eq:diff}) and (\ref{eq:fscat}) to
compute ${dv_{\rm sh}}/{dt}$.

\noindent
{\bf 3.} Over an infinitesimal time step $\Delta t$ a new shell
position and velocity are then generated as $r_{\rm sh}(t+\Delta t)=
r_{\rm sh}(t) + v_{\rm sh}(t)\Delta t$ and $v_{\rm sh}(t+\Delta
t)=v_{\rm sh}(t)+({dv_{\rm sh}}/{dt})\Delta t$.  At the new position,
the shell's column density reduces to $N_{\rm HI}=N_{\rm HI,i}(r_{\rm
sh}/r_{\rm sh,i})^{-2}$, and we return to step ${\bf 2.}$

Our fiducial model has the following parameters.  We adopt
$\sigma=150$ km s$^{-1}$, corresponding to the circular virial
velocity of a dark matter halo of mass $M=10^{11}M_{\odot}$ at $z=5.7$
as appropriate for a typical host of known Ly$\alpha$ emitting
galaxies \citep[e.g.][]{igm}. We also adopt $L_{\alpha}=10^{43}$ erg
s$^{-1}$, corresponding to the typical observed luminosity of
Ly$\alpha$ emitting galaxies \citep[e.g.][]{Ouchi07}. This
Ly$\alpha$ luminosity corresponds to a star formation rate of $\sim 5
M_{\odot}$ yr$^{-1}$ for a Salpeter IMF and a gas metallicity
$Z=0.05Z_{\odot}$ \citep[][for $Z=Z_{\odot}$ this star formation rate
is higher by a factor of 2]{S03}. For comparison, the total kinetic
luminosity in supernova ejecta is $L_{\rm mech}\sim 2 \times
10^{42}(E_{\rm mech}/10^{51}\hs{\rm erg})(\mathcal{N}_{\rm
SN}/10^{-2})$ erg s$^{-1}$, where $\mathcal{N}_{\rm SN}$ is the
supernova rate per unit star formation rate, and  $E_{\rm mech}$ is
the total kinetic energy in ejecta per supernova explosion
\citep[e.g.][and references therein]{Murray05}.

For the shell, we assume $N_{\rm HI,i}=10^{20}$ cm$^{-2}$, $r_{\rm
sh,i}=1.0$ kpc (see \S\ref{sec:intro} and Verhamme et al. 2008) and
$v_{\rm sh,i}=50$ km s$^{-1}$. We find that the final shell speed is
not very sensitive to $v_{\rm sh,i}$. The above shell parameters imply
a shell mass of $m_{\rm sh}\sim 10^7$ M$_{\odot}$. The parameters of
the fiducial model are summarized in Table~\ref{table:param}.

\begin{table}
\centering
\caption{Parameters of Models used in Figure~\ref{fig:speed}.}
\begin{tabular}{c c c c c c}
\hline\hline \# &$\sigma$ & $L_{\alpha}$  & $r_{\rm sh,i}$ & $v_{\rm
sh,i}$ & $N_{\rm HI,i}$\\ &(km s$^{-1}$) &(erg s$^{-1}$) & (kpc)  &
(km s$^{-1}$) & (cm$^{-2}$)\\ \hline 1.& $150$ &$10^{43}$ & $1.0$ &
$50$ & $10^{20}$ \\ 2.& $150$ &$10^{43}$ & $1.0$ & $150$ & $10^{20}$
\\ 3.& $150$ &$10^{43}$ & $1.0$ & $50$ & $10^{19}$ \\ \hline\hline
\end{tabular}
\label{table:param}
\end{table}

The time evolution of the shell speed in the fiducial model is plotted
as the {\it solid line} in the {\it left panel} of
Figure~\ref{fig:speed}, which shows that Ly$\alpha$ scattering
accelerates the shell to $\sim 100$ km s$^{-1}$ after 10 Myr,
consistently with the crude estimate given in
\S~\ref{sec:intro}. After 100 Myr, the shell speed reaches $\sim 230$
km s$^{-1}$. The acceleration of the decreases with time because
$f_{\rm scat}$ decreases with time. As the shell accelerates and
expands, the fraction of Ly$\alpha$ photons that are scattered in the
shell and the associated momentum transfer rate to the shell,
decrease. Other lines in this panel represent models in which one of
the model parameters was changed.

The {\it red dashed line} shows the shell evolution for a model in
which the initial shell velocity was increased to $v_{\rm sh,i}=150$
km s$^{-1}$. The difference in shell speed between this case and the
fiducial model decreases with time. In the model represented by the
{\it green dot-dashed line} the shell mass is reduced to $m_{\rm
sh}=10^6M_{\odot}$ by reducing $N_{\rm HI,i}$. The shell speed
evolves much faster, and reaches $200$ km s$^{-1}$ after $\sim 5$ Myr.
The final shell speed is $\sim 400$ km s$^{-1}$. The {\it central
(right) panel} shows the shell velocity as a function of its radius
(column density). In the fiducial model (and those in which one
parameter was changed), the shell started out at $r=1$ kpc with a
relatively low column density.  The model represented by the {\it blue
dotted lines} is discussed in \S~\ref{sec:modelpar}.

In all the above models, Ly$\alpha$ radiation pressure accelerates the
shell up to a few hundreds of km s$^{-1}$. For comparison, the
escape velocity from a halo with $v_{\rm circ}=150$ km s$^{-1}$ (which
corresponds to a mass of $3\times 10^{11}[(1+z)/4]^{-3/2}M_{\odot}$,
e.g. Barkana \& Loeb 2001) for a shell starting at $r=1$ kpc is
$v_{\rm esc}\sim 440$ km s$^{-1}$ (at $z=3$, with a very weak redshift
dependence, see \S~\ref{sec:modelpar}). Therefore, it is reasonable to
claim that shells that are driven by Ly$\alpha$ radiation pressure may
escape from massive dark matter halos. Next we consider whether this
mechanism could operate in known observed high-redshift LBGs that have
measured values of $L_{\alpha}$, $N_{\rm HI}$ and $v_{\rm sh}$.

\subsection{The Time Evolution of the Shell Speed in Known Galaxies}
\label{sec:modelpar}

In real galaxies, Ly$\alpha$ radiation is not the only process that
determines the shell kinematics, and Equation~(\ref{eq:diff}) modifies
to
\begin{equation}
m_{\rm sh}\frac{dv_{\rm sh}}{dt}=f_{\rm scat}(v_{\rm sh},N_{\rm
HI})\frac{L_{\alpha}}{c}-F_{\rm grav}(r)+4\pi r^2 \Delta P,
\label{eq:diff2}
\end{equation} 
where $F_{\rm grav}(r)$ is the gravitational force on the shell, and
$\Delta P=P_{\rm int}-P_{\rm ext}$ with $P=\rho c_{\rm S}^2$ being the
pressure of the medium inside (``int'') or outside (``ext'') of the
shell \citep{E82}. Here, $\rho$ is the density of the medium, and
$c_{\rm S}$ is its sound speed. The second and third terms on the
right hand side of Eq.~(\ref{eq:diff2}) require assumptions about the
unknown shape of the gravitational potential near the star forming
region and the local properties of the surrounding ISM.

To first order, gravity can be ignored in those galaxies for which the
calculated terminal shell speed exceeds the escape velocity of the
dark matter halo. More specifically, gravity can be ignored if the
Ly$\alpha$ momentum transfer rate exceeds the gravitational force,
${L_{\alpha}}/{c}\gsim GM(<r_{\rm sh})m_{\rm sh}/r_{\rm sh}^2$, where
$M(<r_{\rm sh})$ is the total mass enclosed by the
supershell. Quantitatively, Ly$\alpha$ radiation pressure exceeds
gravity when the Ly$\alpha$ luminosity exceeds
\begin{equation}
L_{\alpha}\gsim 10^{43}\hs{\rm erg}\hs {\rm
s}^{-1}\Big{(}\frac{M(<r_{\rm sh})}{10^8
M_{\odot}}\Big{)}\Big{(}\frac{m_{\rm
sh}}{10^6M_{\odot}}\Big{)}\Big{(}\frac{r_{\rm sh}}{0.1\hs{\rm
kpc}}\Big{)}^{-2}.
\end{equation} 
If the radial density profile can be modeled as an isothermal sphere,
then $M(<r_{\rm sh})\sim M_{\rm tot}(r_{\rm sh}/{r_{\rm vir}})$ and
the required luminosity increases at smaller shell radii. However, if
the shell starts at a small radius (i.e. $r_{\rm sh,i}=0.01$ kpc),
then for a fixed shell mass the column density is high (e.g. $N_{\rm
HI,i}\gg 10^{22}$ cm$^{-2}$). This, in combination with the small
initial velocity of the shell, implies that the Ly$\alpha$ photons are
efficiently trapped and the impact of radiation pressure is bigger
than estimated above (see Fig.~\ref{fig:fscat} and Paper I). In Paper
I we showed that in this regime, a simple order-of-magnitude estimate
for the importance of radiation pressure is obtained based on energy
considerations \citep[e.g.][]{Cox85,Bithell90,Oh02}. Initially, the
total gravitational binding energy of the shell is $|U_{\rm
sh}|=GM(<r_{\rm sh,i})m_{\rm sh}/r_{\rm sh,i}$. The energy in the
Ly$\alpha$ radiation field is $U_{\alpha}=L_{\alpha}t_{\rm trap}$,
where $t_{\rm trap}$ is the typical trapping time of a Ly$\alpha$
photon in the neutral shell. The trapping time scales with the shell
optical depth at the Ly$\alpha$ line center, $\tau_0$, through the
relation $t_{\rm trap}\sim 15(\tau_0/10^{5.5})r_{\rm sh,i}/c$
\citep{Adams75,Bo79}. Radiation pressure unbinds the gas when
$U_{\alpha}>|U_{\rm sh}|$. We find this to be the case for all
galaxies discussed below (except the first model of FDF 1267). Also
$U_{\alpha}\gg |U_{\rm sh}|$ for those galaxies in which the
calculated terminal wind speed exceeds the escape velocity of the
halo, which confirms that ignoring gravity is justified best for those
galaxies in which the Ly$\alpha$ radiation pressure can accelerate the
shell to a speed that exceeds the escape velocity of the host dark
matter halo.

Since we investigate the properties of Ly$\alpha$-driven supershells,
we assume that $\Delta P$ is less than the Ly$\alpha$ radiation
pressure (this assumption is motivated by the possibility that
Ly$\alpha$ radiation pressure provides the dominant source of pressure
on neutral hydrogen gas in the ISM, see \S~\ref{sec:intro} and Paper
I).  Hence, we solve Eq.~\ref{eq:diff2} under the assumption that the
second and third terms on its right hand side can be ignored. We
initiate a shell at rest ($v_{\rm sh,i}=0$) at a very small radius
($r_{\rm sh,i}=0.01$ kpc, and not at the origin to avoid a divergent
$N_{\rm HI,i}\rightarrow \infty$). We get constraints on the mass, age
and radius of the supershell by requiring that it reaches its observed
expansion velocity at its observed column density (see
\S~\ref{sec:cB-58} for an example).

Our results are summarized in Table~\ref{table:predict}. The first
column contains the name of the galaxy. The second column contains the
HI column density and the third contains the shell speed as derived
from fitting to the observed Ly$\alpha$ line profile
\citep{SV08,Verhamme08}. The fourth column contains the {\it
intrinsic} Ly$\alpha$ luminosity, which was inferred from the UV-based
star formation rate, and the {\it intrinsic} Ly$\alpha$ equivalent
width derived by Schaerer \& Verhamme (2008) and \citet{Verhamme08},
as\footnote{This implies that we convert a rest-frame
UV luminosity density into a Ly$\alpha$ luminosity using only the
intrinsic EW given by \citet{Verhamme08}. The actual star formation
rates that are associated with these UV luminosity densities are
irrelevant.} $L_{\alpha}\sim ({\rm SFR/M_\odot/yr}) \times ({\rm
EW}_{\rm int}/70\hs {\rm \AA})\times 10^{42}$ erg s$^{-1}$
\citep[e.g.][]{DW07}.  The radius ($r_{\rm sh}$), angular scale on the
sky ($\theta_{\alpha}$), mass ($m_{\rm sh}$), and age ($t_{\rm sh}$)
of the supershell are given in the fifth, sixth, seventh, and eighth
columns, respectively. The ninth column contains the energy of the
supershell, which is given by $E_{\rm sh}=\frac{1}{2}m_{\rm
sh}v^2_{\rm sh}$. The tenth column gives the asymptotic (terminal)
shell speed, to be compared with the escape velocity of the galaxy in
the eleventh column. For an isothermal sphere, the escape velocity
relates to the circular velocity as $v_{\rm esc}=\sqrt{2}v_{\rm
circ}\sqrt{{\rm ln}[r_{\rm vir}/r_{\rm sh,i}]}$ (see
Eqs.~(2-25),(4-117) \& (4-123) of Binney \& Tremaine 1987, and
Eqs.~(24) \& (25) of Barkana \& Loeb 2001). The circular velocity
$v_{\rm circ}$ was obtained from $v_{\rm circ}$=FWHM/2.35, in which
'FWHM' is the intrinsic Full Width at Half Maximum of the Ly$\alpha$
line quoted by Verhamme et al. (2008, see e.g. Dijkstra et al. 2007
for a more extended discussion on this relation).
\begin{table*}
\centering
\caption{Predicted physical sizes of expanding HI supershells under
the assumption that {\it their expansion was driven predominantly by
Ly$\alpha$ radiation pressure}.}
\begin{tabular}{c c c c c c c c c c c c c}
\hline\hline Galaxy & $N_{\rm HI}$ & $v_{\rm sh}$ & $L_{\alpha}$ &
$r_{\rm sh}$ & $\theta_{\alpha}$ & $m_{\rm sh}$ & $t_{\rm sh}$ &
$E_{\rm sh}$& $v_{\rm term}$ & $v_{\rm esc}$ & f &$\frac{\dot{M}}{{\rm
SFR}}$\\ name &($10^{20}$/cm$^{2}$) & (km/s) & ($ 10^{43}$ $\frac{{\rm
erg}}{{\rm s}}$) & (kpc) & (\arcsec) &($10^6 M_{\odot}$) & (Myr) &
($10^{54}$ erg) & (km/s)& (km/s) & \\ \hline cB-58& 7.5& 200.& 6.&
0.22& 0.03& 3.3& 1.9 & 1.3& 370& 134& 1.0 & 0.24\\ \hline
\multicolumn{12}{c}{FORS Deep Field Galaxies observed by Tapken et al
(2007). Parameters taken from Verhamme et al (2008).}\\ \hline 1267&
0.2& 50.& 0.3& 5.5& 0.69& 58& 200& 1.4& 110.& 544& 0.85 & 1.4\\ 1267&
3& 300.& 1.& 0.03& 0.004& 0.03& 0.14& 0.02& 544& 170& 1.0 & 0.7\\ 1337
& 5& 200.& 2.5& 0.12& 0.016& 0.7& 1.05 & 0.27& 437& 170& 0.89 & 0.1\\
2384 & 0.3& 150.& 5.& 7.4& 1.0& 157& 89& 35& 164& 170& 0.92 & 0.4\\
3389 & 0.2& 150.& 1.2& 2.5& 0.37& 12& 29.5& 2.7& 210& 263& 0.87 &
0.2\\ 4454 & 0.2& 150.& 0.28& 0.63& 0.8& 0.08& 7.4& 0.2& 211& 256&
0.92 & 0.3\\ 5215 & 7& 400.& 4.4& 0.03& 0.004& 0.06& 0.1& 0.1& 780&
170& 0.65 & 0.1\\ 5550 & 5& 200.& 4.2& 0.21& 0.03& 2& 1.9& 0.84& 430&
170& 0.93 & 0.1\\ 5812 & 0.2& 150.& 2.0& 4.3& 0.67& 35& 50& 8& 196&
165& 0.90 & 0.7\\ 6557 & 0.4& 150.& 1.4& 1.7& 0.26& 11& 22& 2.5& 223.&
166& 1.0 & 0.2\\ 4691 & 0.8 & 10 & 18 &230 &30 &$4 \times 10^5$&
$4\times 10^{4}$& 400 &- &1936 &0.83 & 3.9 \\ 7539 & 5.0 &25 &42 &15 &
2.0& $10^4$ & $10^3$ &67 &- &170 & 1.0 &1.9 \\\hline\hline
\end{tabular}
\label{table:predict}
\end{table*}
The twelfth column contains a fudge factor $f$ which is defined as
$f\equiv r_{\rm sh}/(\frac{1}{2}\tilde{a}_{\rm sh}\tilde{t}^2_{\rm
sh})$, where $\tilde{a}_{\rm sh}\equiv{L_{\alpha}}/{cm_{\rm sh}}$, and
$\tilde{t}_{\rm sh}\equiv 2r_{\rm sh}/v_{\rm sh}$ (i.e. $f=1$ for
$f_{\rm scat}=1$, and $f<1$ when $f_{\rm scat}<1$). The radius of the
shell can then be expressed as a function of the observed quantities
$L_{\alpha}$, $N_{\rm HI}$, and $v_{\rm sh}$ as
\begin{equation}
r_{\rm sh}=\frac{f}{2\pi \mu m_p c}\frac{L_{\alpha}}{N_{\rm HI}v_{\rm
sh}^2}
\label{eq:rsh}
\end{equation} 

Finally, the thirteenth column contains the ratio between the mass
outflow rate in the supershell given by $\dot{M}$= $6.0\times (r_{\rm
sh}/1 \hs {\rm kpc })$$(N_{\rm HI}/10^{20}\hs{\rm cm}^{-2})(v_{\rm
sh}/200\hs{\rm km}\hs{\rm s}^{-1})$ $M_{\odot}$ yr$^{-1}$
\citep{Verhamme08}, and the UV-derived star formation rate (in
$M_{\odot}$ yr$^{-1}$).

Throughout our calculations of $f_{\rm scat}$ we have used the
$b$--parameter similarly to \citet{SV08} and \citet{Verhamme08}
instead of $v_{\rm th}$, i.e. we substituted $\Delta \nu_D
=b\nu_{\alpha}/c$ in equation~(\ref{eq:fscat}). Below, we describe
individual cases in more detail.

\subsubsection{MS1512-cB58}
\label{sec:cB-58}
\citet{Pettini02} have found that the lensed Lyman Break galaxy (LBG)
MS1512-cB58 is surrounded by an outflowing shell of gas with a column
density $N_{\rm HI}=7.5\times 10^{20}$ cm$^{-2}$, which is expanding
at a speed of $v_{\rm sh}=200$ km s$^{-1}$ \citep[also
see][]{SV08}. The star formation rate inside the LBG, $\sim 40
M_{\odot}$ yr$^{-1}$, translates to an intrinsic Ly$\alpha$ luminosity
of $\sim 4$--$8\times 10^{43}$ erg~s$^{-1}$ (with the uncertainty due
to the unknown metallicity of the gas, see Schaerer 2003). We adopt
the central value of $L_{\alpha}=6\times 10^{43}$ erg~s$^{-1}$.

The {\it blue dotted line} in Figure~\ref{fig:speed} depicts the time
evolution of a shell with $m_{\rm sh}=3\times 10^6M_{\odot}$ and the
above parameters of MS1512-cB58. The plot shows that the observed
shell velocity is reached at $t\sim 1.9$ Myr, when the shell reaches a
radius $r=0.21$ kpc and has an observed column density of $N_{\rm
HI}=7.5 \times 10^{20}$ cm$^{-2}$. More massive shells would reach the
observed speed at later times, larger radii, and at a lower shell
column density. The plot also shows that Ly$\alpha$ pressure
accelerates the shell to $\sim 360$ km s$^{-1}$ after 10 Myr, at which
point the column density has declined to $\sim 4\times10^{18}$
cm$^{-2}$.

Lastly, we point that the HI shell in  MS1512-cB58 is known to
contain dust, with an estimated extinction of E$(B-V)\sim 0.3$
\citep[][and references therein]{SV08}, which implies an approximate
optical depth at $1216$ \AA\hs of $\tau_{\rm D}\sim 1.8-3.3$
\citep[e.g.][]{Verhamme08}. We argue in \S~\ref{sec:conc} that despite
the presence of this dust, the Ly$\alpha$ radiation pressure can still
be substantial (and possibly dominant over continuum radiation
pressure).

\subsubsection{Ly$\alpha$ Emitters in the FORS Deep Field}
\label{sec:FORS}

\citet{Verhamme08} reproduced the observed Ly$\alpha$ line profiles of
11 LBGs from the FORS Deep Field at $2.8 \lsim z \lsim 5$ observed by
\citet{Tapken07} based on a simple model in which the Ly$\alpha$
photons emitted by the LBGs backscatter off a single spherical
outflowing shell with $2\times 10^{19}$cm$^{-2}\lsim N_{\rm HI} \lsim
7\times 10^{20}$ cm$^{-2}$.

Table~\ref{table:predict} contains two entries for FDF 1267 because
two different models were found to reproduce the observed line
profile. We have not included galaxies FDF 4691 and FDF 7539 in the
table because these galaxies contained almost static supershells,
which cannot be reproduced by our approach, possibly because the
shells in these galaxies were slowed down by gravity.

For several galaxies our model can be ruled out: our predicted
$r_{\rm sh}$ for FDF 4691 exceeds the size of the largest observed
Ly$\alpha$ emitting structure in our Universe, which is $\sim 150$ kpc
in diameter \citep{Steidel00}. Furthermore, we find ages of $t_{\rm
sh}\sim 0.1$ Myr for two galaxies (FDF 5215 and FDF 1267). While this
is not physically impossible, it appears unlikely that we have caught
star forming regions that early in their evolution. Indeed, these
predicted ages fall well below the plausible age range that was
derived for local supershells. Lastly, we predict a radius $r_{\rm
sh}=15$ kpc for FDF 7531. Our model therefore predicts that existing
observations should have resolved this galaxy as a spatially extended
Ly$\alpha$ source (which is not observed). 

For most other galaxies, $r_{\rm sh}\sim 0.1$--$10$ kpc, $t_{\rm
sh }\sim 1$--$200$ Myr, and $E_{\rm sh}\sim 10^{53}$--$10^{55}$
ergs. For comparison, \citet{M02} find $r_{\rm sh}\sim 0.07$--$0.7$
kpc, $t_{\rm sh}\sim 0.9$--$20$Myr, and $E_{\rm
sh}=0.03$--$5\times10^{53}$ ergs for 19 galactic supershells. Our
largest shells are larger than those of supershells that were observed
in our galaxy, but are comparable in size of the inferred sizes of HI
outflows around local star burst galaxies (Mass-Hesse et
al. 2003). Thus, there is considerable overlap in the physical
properties of observed (extra)galactic supershells and those in
our model, although our model does contain a few supershells that are
significantly larger and more energetic than observed galactic
supershells. This may be because we ignore gravity. Indeed, for these
energetic shells ($E_{\rm sh}\gsim$ a few $10^{54}$ erg), the maximum
shell speed does not exceed the escape velocity of the host dark
matter halo, and so gravity cannot be ignored.

We caution that the escape velocity of the dark matter halo that we
computed are approximate. Our escape velocity was derived using the
relation $v_{\rm esc}=\sqrt{2}v_{\rm circ}\sqrt{{\rm ln}[r_{\rm
vir}/r_{\rm sh,i}]}$, which assumes that the shell climbs out of the
gravitational potential of an isothermal sphere\footnote{The
isothermal sphere model for the inner mass distribution of galaxies is
supported observationally by gravitational lensing studies of
elliptical galaxies \citep[e.g.][]{Winn03,Bo08} and by the observed
flatness of the rotation curves of spiral galaxies
\citep[e.g.][]{B89,SM02}}. Although our calculated $v_{\rm esc}$
depends only weakly on $r_{\rm sh,i}$, we caution that deviations from
our assumed simple model may change the results somewhat.

Interestingly, the fudge factor $f$ is within the range $0.65$--$1.0$
in all models. The small scatter in $f$ arises because all shells
initially have much higher HI column densities than their observed
values. Therefore, for all shells the parameter $f_{\rm scat}=1$
during the early stages of their evolution. Depending on the precise
time evolution of column density and shell speed, $f_{\rm scat}$
eventually drops below unity (for some galaxies this has not happened
yet and $f_{\rm scat}=1$, which in turn implies $f=1$). Additional
scatter may arise from the fact that some shells have predicted size
that are close to the initial shell size assumed for our models.

The small scatter in $f$ implies that one can predict physical
properties of galactic supershells without numerically integrating
equation~(\ref{eq:diff2}).  Instead, one may simply adopt the central
value $f\sim 0.9$ and apply Eq.~(\ref{eq:rsh}) to {\bf } quantities
such as $L_{\alpha}$, $N_{\rm HI}$ and $v_{\rm sh}$ to predict $r_{\rm
sh}$. We also find that the mass outflow rates in supershells reach
$\sim 10$--$100\%$ of the star formation rates in their host galaxies,
consistently with the {\it total} outflow rates that one expects
theoretically \citep[e.g.][]{Erb08}.

\section{Discussion \& Conclusions}
\label{sec:conc}
It is well known that scattering of Ly$\alpha$ photons by neutral
hydrogen in a single outflowing supershell around star-forming
galaxies can naturally explain two observed phenomena: ({\it i}) the
common shift of the Ly$\alpha$ emission line towards the red relative
to other nebular recombination and metal absorption lines, and ({\it
ii}) the asymmetry of the Ly$\alpha$ line with emission extending well
into its red wing.  In this paper we have computed the radiation
pressure that is exerted by this scattered Ly$\alpha$ radiation on the
outflowing shell.

We have shown that for reasonable shell parameters the shell can be
accelerated to a few hundred km s$^{-1}$ after $\lsim 10$ Myr.  The
Ly$\alpha$ acceleration mechanism becomes increasingly inefficient as
the shell speed increases (Fig.~\ref{fig:fw}), because the Doppler
boost of the gas extends increasingly farther into the wing of the
Ly$\alpha$ absorption profile allowing a decreasing fraction of the
Ly$\alpha$ photons to scatter on the outflowing shell.  
Furthermore, for a shell mass that is constant in time, $N_{\rm
HI}\propto r^{-2}$, and we find that for any choice of model
parameters the shell achieves a terminal velocity that depends
mainly\footnote{An alternative way to see why the observed quantities
$N_{\rm HI}$, $v_{\rm sh}$, and $L_{\alpha}$ provide unique
constraints on the shell mass, radius, and age is that these three
unknown quantities relate to the observed quantities via three
equations: {\bf 1. $m_{\rm sh}=4\pi r^2_{\rm sh}N_{\rm HI}\mu m_p$}
and {\bf 2.} $v_{\rm sh}\sim \frac{L_\alpha}{cm_{\rm sh}}t_{\rm sh}$,
and {\bf 3. $r_{\rm sh}\sim \frac{L_\alpha}{2cm_{\rm sh}}t^2_{\rm
sh}$}. We found that Equations {\bf 2.} and {\bf 3.} are accurate to
within $\sim 50\%$.} on its total mass and the Ly$\alpha$ luminosity
of the central source (Fig.~\ref{fig:speed}).

We have computed the physical properties of expanding supershells that
are likely to be present in specific observed high-redshift
($z=2.7$--$5.0$) galaxies, under the assumptions that they are driven
predominantly by Ly$\alpha$ radiation pressure and that their mass
remains constant in time. We predict supershell radii that lie in the
range $r_{\rm sh}=0.1$--$10$ kpc, ages in the range $t_{\rm sh
}=1$--$100$ Myr, and energies in the range $E_{\rm
sh}=10^{53}$--$10^{55}$ ergs, in broad agreement with the properties
of local galactic supershells. We derive mass outflow rates in the
supershells that reach $\sim 10$--$100\%$ of the star formation rates
in the host galaxies, in agreement with the {\it total} outflow rates
that one expects theoretically \citep[e.g.][]{Erb08}.

We have found that in all models the radius of the supershell is
determined uniquely by parameters such as the {\it intrinsic}
Ly$\alpha$ luminosity of the host galaxy (which may be inferred
from the observed Ly$\alpha$ line shape), $L_{\alpha}$, the total
column density of HI in the supershell, $N_{\rm HI}$, and the shell
velocity $v_{\rm sh}$,
\begin{equation}
\left(\frac{r_{\rm sh}}{{\rm kpc}}\right)\sim
0.2\Big{(}\frac{f}{0.9}\Big{)}\Big{(}\frac{L_{\alpha}}{10^{43}\hs
  \frac{{\rm erg}}{{\rm s}}}\Big{)}\Big{(}\frac{10^{20}/{\rm
    cm}^{2}}{N_{\rm HI}}\Big{)}\Big{(}\frac{200\hs\frac{{\rm km}}{{\rm
      s}}}{v_{\rm sh}}\Big{)}^{2}.
\label{eq:rshell}
\end{equation}
Here $f$ is a fudge factor that lies in the range $f=1.0$--$1.3$. Our
models have ignored gravity and the pressure of exerted by the
surrounding interstellar medium. {\it This relation holds up only for
supershells that are driven predominantly by Ly$\alpha$ radiation
pressure}. This relation is most accurate for those galaxies in which
the shells are accelerated to velocities that exceed $v_{\rm esc}$. We
found this to be the case in 5 out of 9 (55\%, see
Table~\ref{table:predict}, ignoring the galaxies for which our model
was ruled out, i.e. FDF 1267b, FDF 5215, FDF4691, and FDF 7539) of
known galaxies. Hence, these shells could contribute to the enrichment
of the intergalactic medium.

In other cases, gravity (and external pressure) will tend to reduce
the value of the fudge factor $f$. Also, if neutral hydrogen gas only
makes up a fraction $f_{\rm HI}$ of the total mass of the shell (see
\S~\ref{sec:intro}), then our predicted value of $r_{\rm sh}$ should
be lowered by a factor of $f_{\rm HI}$.

\begin{figure}
\vbox{\centerline{\epsfig{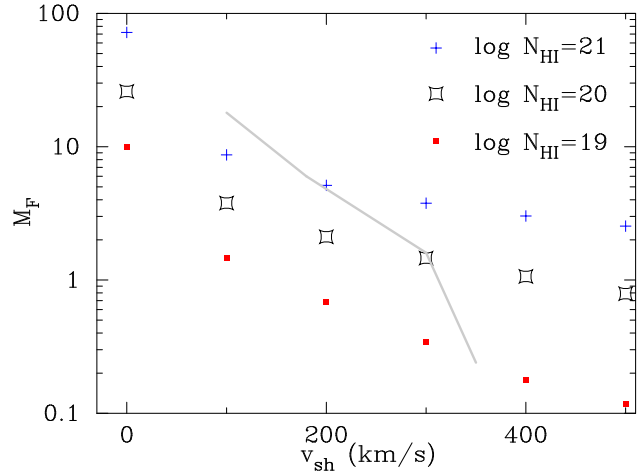}}}
\caption[]{When Ly$\alpha$ trapping is properly accounted for, the
total radiation force in Eq.~\ref{eq:diff} should be modified as
$f_{\rm scat}(v_{\rm sh},N_{\rm HI}) \rightarrow M_{F}(v_{\rm
sh},N_{\rm HI})$. Here, we plot $M_F$ (a 'force-multiplication'
factor) as a function of the expansion velocity of the HI shell,
$v_{\rm sh}$, for three values of HI column density. $M_F$ was
calculated by performing Monte-Carlo Ly$\alpha$ radiative transfer
calculation (described in Paper I), and under the assumption that the
HI shell surrounds an empty cavity. $M_F$ increases with increasing
$N_{\rm HI}$ and decreasing $v_{\rm sh}$. The {\it grey line} is the
trajectory in the $N_{\rm HI}-v_{\rm sh}$ plane of the model of the
supershell in cB-58 (\S~\ref{sec:cB-58}). Clearly, especially at early
times we have underestimated the radiation pressure force
significantly, which renders our calculations conservative.}
\label{fig:fscat}
\end{figure}

On the other hand, as was already mentioned in \S~\ref{sec:result},
our conservative calculations have completely ignored the fact that
'trapping' of Ly$\alpha$ radiation by the optically thick supershell
may significantly boost the Ly$\alpha$ radiation pressure compared to
what was used in this paper. Specifically, it was shown in Paper I
that when this trapping is properly accounted for, the total radiation
force in Eq.~\ref{eq:diff} should include $M_{F}(v_{\rm sh},N_{\rm
HI})$ in place of $f_{\rm scat}(v_{\rm sh},N_{\rm HI})$. Here, $M_F$
is 'force-multiplication' factor\footnote{This term derives from the
(time-dependent) force-multiplication function $M(t)$ that was
introduced by \citet{Castor75}, as $F_{\rm rad}\equiv
M(t)\frac{\tau_eL_{\rm bol}}{c}$. Here, $F_{\rm rad}$ is the total
force that radiation exerts on a medium, $\tau_e$ is the total optical
depth to electron scattering through this medium. The function $M(t)$
arises because of the contribution of numerous metal absorption lines
to the medium's opacity, and can be as large as $M_{\rm max}(t)\sim
10^3$ in the atmospheres of O-stars \citep{Castor75}.} that depends
both the column density of HI and the velocity of the supershell. In
Paper I, $M_F$ was computed for a range of $N_{\rm HI}$ and $v_{\rm
sh}$ values, by performing Monte-Carlo Ly$\alpha$ radiative transfer
calculation for each combination of $N_{\rm HI}$ and $v_{\rm sh}$. In
these calculations, the HI shell surrounds an empty cavity. The
results of these calculations are shown in Fig.~\ref{fig:fscat}. As
evident from the plot, $M_F$ greatly exceeds unity for low shell
velocities and large HI column densities. Overplotted as the {\it grey
line} is the trajectory in the $N_{\rm HI}-v_{\rm sh}$ plane of the
supershell in cB-58 (\S~\ref{sec:cB-58}). At early times, $M_F> 10$,
and we may have underestimated the radiation pressure force
significantly (as was mentioned in \S~\ref{sec:result}, the actual
force multiplication factor may be less than $M_F$, if gas and dust
interior to the supershell prevents photons from repeatedly 'bouncing'
back an forth between opposite sides of the HI shell). Ly$\alpha$
trapping could boost our predicted value of $r_{\rm sh}$ by a factor
as large as $\langle M_F \rangle$, where $\langle M_F \rangle$ the
value of $M_F$, averaged over the expansion history of the shell.

So far, we have assumed HI to be the only source of opacity in the
supershell. However, supershells may also contain dust which provides
an additional source of opacity; for example, in the models of
Verhamme et al (2008), the optical depth through dust at
$\lambda=1216$ \AA\hs is $\tau_{\rm D}=0.0$--$2.0$. At a given HI
column density, we have therefore systematically underestimated
$f_{\rm scat}$, (absorption of a Ly$\alpha$ photon by dust also
results in a momentum transfer of magnitude $h\nu_{\alpha}/c$), which
renders our calculations conservative. Furthermore, dusty supershells
could also absorb significant amounts of continuum radiation, boosting
the total momentum transfer rate even further. If the dust
resides interior to the HI shell, then the Ly$\alpha$ flux that
impinges upon the shell is reduced by
e$^{-\tau_d(\lambda_{\alpha})}$. In this case, the Ly$\alpha$ (as well
as the continuum) radiation pressure is smaller than computed in this
paper. However, the coupling between gas and dust makes it very
unlikely that (Ly$\alpha$) radiation pressure sweeps up the gas
surrounding a star forming region, while keeping the dust in place
\citep[see][for more discussion on this]{Murray05}.

Of course, the existence of dust inside the HI shell reduces its
ability to 'trap' Ly$\alpha$ photons, which reduces the value of $M_F$
compared to that shown in Figure~\ref{fig:fscat}. Despite this
reduction, $M_F$ can still greatly exceed unity, because at low shell
expansion velocities the majority of Ly$\alpha$ photons do no
penetrate deeply into the HI shell upon their first entry. Instead,
the photons mostly scatter near the surface of the shell, and only
'see' a fraction of the dust opacity (indeed, for this reason
Ly$\alpha$ radiation can escape from a dusty two-phased ISM, see
Neufeld 1991 and Hansen \& Oh 2006). In other words, a relative small
inner portion the HI shell can effectively trap Ly$\alpha$ photons. We
have verified this statement with Monte-Carlo radiative transfer
calculations that included dust at the levels inferred by
\citet{Verhamme08}: we typically found $M_F$ to be reduced by a factor
of up to a few. In other words, even for dusty shells it is possible
that $M_F\gg 1$.

Therefore, the pressure exerted by Ly$\alpha$ radiation alone can
exceed the maximum possible pressure exerted by continuum radiation
(also see Paper I), which can drive the large gas masses (as large as
a few $10^{10-11}M_{\odot}$) out of galaxies in a galactic {\it
superwind} \citep{Murray05}. For comparison, in our model the
Ly$\alpha$ photons transfer their momentum onto the expanding
supershell, which contains significantly less mass ($m_{\rm sh}\lsim
10^8 M_{\odot}$). This implies that in principle, both mechanisms
could operate simultaneously, and that the neutral HI supershells may
be accelerated to higher velocities than the bulk of the ejected
gas. Since in our model Ly$\alpha$ radiation pressure operates only on
a fraction of the gas, it does not provide a self-regulating mechanism
for star formation and black hole growth as in
\citet{Murray05}. However, it does provide a new way of enriching the
intergalactic medium with metals.

We note that quasars can have Ly$\alpha$
luminosities\footnote{Empirically, the Ly$\alpha$ luminosity of
quasars is related to their B-band luminosity as $L_{\alpha}\sim
0.7L_{\rm B}$ (Dijkstra \& Wyithe, 2006). High-redshift quasars have
been observed with $L_{\rm B}\gsim 10^{13}L_{{\rm B},\odot}=4\times
10^{45}$ erg s$^{-1}$ (see e.g Figure~1 of Dijkstra \& Wyithe, 2006).}
that reach $L_{\alpha}=10^{46}$ erg s$^{-1}$ \citep[][]{Fan06}. In
principle, this Ly$\alpha$ luminosity could transfer a significant
amount of momentum onto neutral gas in its proximity. Because the
Ly$\alpha$ emission line of quasars is typically very broad ($\sim$
few thousand km s$^{-1}$), $f_{\rm scat}$ is significantly less than
unity for the column densities considered in the paper (see
Fig.~\ref{fig:fw}). Furthermore, it is unclear whether gas clouds can
remain neutral in close proximity to the quasar for an extended period
of time. Even if this is the case, it remains to be determined whether
the Ly$\alpha$ radiation pressure can compete with the continuum
radiation pressure. The importance of Ly$\alpha$ radiation pressure
near quasars therefore remains an open issue.

To conclude, observations indicate that it is the kinematics of HI gas
surrounding star forming regions that mostly determines the observed
properties of the Ly$\alpha$ radiation \citep[as opposed to dust
content, e.g.][]{Kunth98,Atek08,Hayes08,Ostlin08}. This work suggests
that -at least in some cases- the pressure exerted by the Ly$\alpha$
photons themselves may be important in determining the kinematics of
HI gas. This is appealing for two reasons: ({\it i}) the mechanism
operates irrespective of the dust content of the HI supershells. This
mechanism may therefore operate also in galaxies of primordial
composition at high redshift (especially since the total Ly$\alpha$
luminosity per unit star formation rate is higher, e.g. Schaerer
2003), and ({\it ii}) the shape and velocity offset of the observed
Ly$\alpha$ emission strongly suggest that momentum transfer from
Ly$\alpha$ photons to HI gas is actually observed to occur.
 
{\bf Acknowledgments} This work is supported by in part by NASA grant
NNX08AL43G, by FQXi, and by Harvard University funds. We thank an anonymous referee for constructive comments that improved the presentation of this paper.

\label{lastpage}
\end{document}